# EXAMPLE OF A NON-SMOOTH HOPF BIFURCATION IN AN AERO-ELASTIC SYSTEM

Luca Magri and Ugo Galvanetto

E-mail: luc.magri@gmail.com; ugo.galvanetto@unipd.it

**Abstract** We investigate a typical aerofoil section under dynamic stall conditions, the structural model is linear and the aerodynamic loading is represented by the Leishman-Beddoes semi-empirical dynamic stall model. The loads given by this model are non-linear and non-smooth, therefore we have integrated the equation of motion using a Runge-Kutta-Fehlberg algorithm equipped with event detection. The main focus of the paper is on the interaction between the Hopf bifurcation typical of aero-elastic systems, which causes flutter oscillations, and the discontinuous definition of the stall model. The paper shows how the non-smooth definition of the dynamic stall model can generate a non-smooth Hopf bifurcation. The mechanisms for the appearance of limit cycle attractors are described by using standard tools of the theory of dynamical systems such as phase plots and bifurcation diagrams.

## 1. - INTRODUCTION

Dynamic stall is a complex unsteady flow phenomenon characterized by the formation, convection and shedding of vortices on the suction side of an aerofoil [1-3]. It may induce oscillations in turbo-machinery compressor blades, helicopter blades, wind turbines and other streamlined structures operating at high angles of incidence. The relevant aerodynamic loads on the aerofoil can be significantly different from the loads corresponding to the same angle of incidence in the static case. A thorough description of dynamic stall with particular emphasis on modelling can be found in reference [4]. In the present paper only a brief description of the most important aspects of flow development about the aerofoil will be given to describe the effects of dynamic stall on the aerodynamic loading.

Figure 1 illustrates the aerodynamic loading due to dynamic stall for a pure pitching oscillation of an aerofoil. The main steps during a dynamic stall cycle are indicated by numbers on the curves in this figure. At point 1 the dynamic lift exceeds maximum static lift when its angle of incidence is close to the static stall angle. A vortex is then formed at the leading edge at point 2 which also marks the beginning of the moment stall. The vortex, attached to the airfoil, grows in strength, is convected along the suction side, reaches the trailing edge and begins its detachment at point 3, which also corresponds to the lift stall. Finally, when the angle of incidence is low enough again, flow reattachment begins at point 4.

The presence of an attached vortex on the suction side of the aerofoil generates a higher lift whereas the detachment of this vortex results in a substantial lift reduction.

Several semi-empirical dynamical stall models have been proposed in the literature [5-10]. In the present paper we have chosen the state-space formulation of the Leishman-Beddoes model (LB model) [11-13] because this model and its many modifications [14, 15] are widely used in engineering applications.

Research on the dynamics of non-smooth dynamical systems is a fast growing field [16-20]. However the complexity of the LB model and its relatively large size prevent us from using some of the powerful theories currently under development for the analysis of such systems. We have chosen the structural model to be linear so as to ensure that the non-linear behaviour is exclusively due to the aerodynamics loading represented by the LB model.

The present papers differs from others on a similar topic [21-25] because it is mainly concerned with the effects of the interaction between the discontinuous definition of an aero-elastic system and its main Hopf bifurcation.

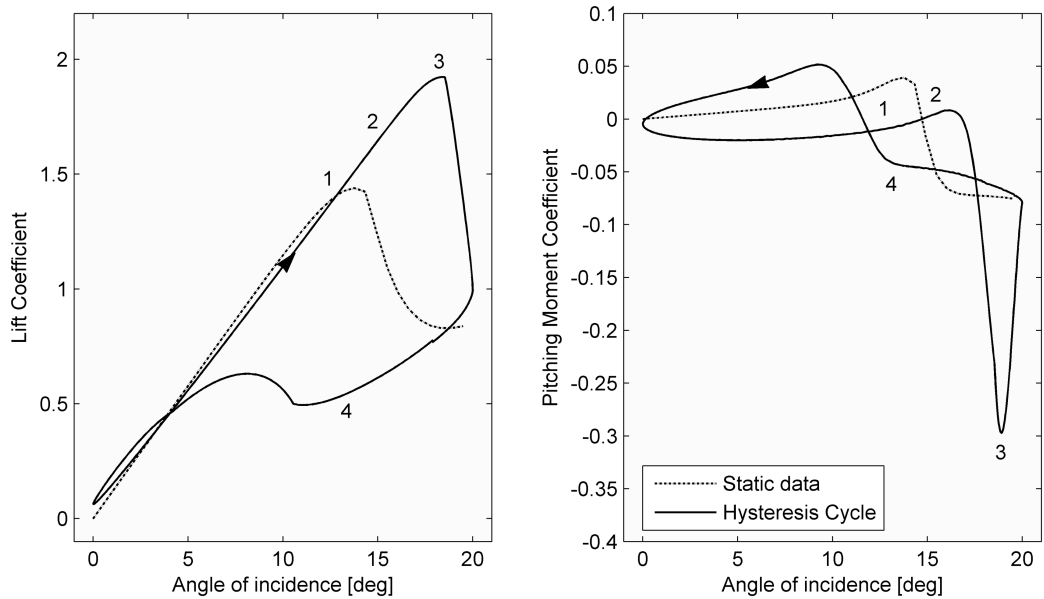

**Figure 1**: Static (dashed line) and dynamic (solid line) aerodynamic forces and moments of an aerofoil in pure pitching oscillation. The numbers refer to key dynamic stall events described in the text.

The paper is organised as follows. Section 2 presents the governing equations of motion describing the aero-elastic behaviour of a typical aerofoil section with two degrees of freedom: pitch and plunge. Section 3 describes the LB model and its implementation focussing in particular on the non-smooth definition of its equations. Section 4 presents two examples of Hopf bifurcations the second of which is caused by the non-smoothness of the aero-elastic system. The conclusions summarise the main findings of the paper.

**2. – STRUCTURAL DYNAMICS OF A TYPICAL AEROFOIL SECTION**

The mechanical system under investigation is a two-dimensional aerofoil section immersed in a horizontal flow of undisturbed speed $U$, as shown in fig. 2. The two degrees of freedom are: displacement in the vertical direction (plunge) and rotation in the plane of the figure (pitch). Plunge is denoted by $h$, taken to be positive in the downward direction, and $\alpha$ is the pitch angle about the elastic axis, positive if nose-up. $c$ is the chord length and $b = c/2$ represents the mid-chord length. The elastic axis is located at a distance $a_h b$ from the mid-chord, whereas the mass centre is located at a distance $x_\alpha b$ along the chord from the elastic axis. Both distances, $a_h b$ and $x_\alpha b$, are positive when measured toward the trailing edge of the aerofoil. The non-dimensional coefficient $a_h$ may exclusively assume values in the range between -1 and 1. The aerofoil under investigation is a NACA 0012. The structural behaviour is modelled by means of linear bending and torsional springs and dampers which are attached at the elastic axis of the aerofoil. In this paper we describe the aero-elastic system by using non-dimensional parameters. In fact, it is common practice in aerodynamics to use non-dimensional aerodynamic coefficients instead of dimensional physical quantities such as forces and moments. In two-dimensional flows we use normal and chord force coefficients, $C_N$, $C_C$ or equivalently lift and drag coefficients, $C_L, C_D$ and pitching moment coefficient, $C_M$. Every aerodynamic coefficient has been evaluated about the aerofoil quarter chord. Normal and chord forces act perpendicular and parallel to the airfoil chord respectively, while lift and drag forces act perpendicular and parallel to the free-stream direction. As a consequence of this geometric definition, we recall two well-known aerodynamic relations:

$$C_L = C_N \cos\alpha - C_C \sin\alpha, \qquad C_D = C_N \sin\alpha + C_C \cos\alpha \qquad (1)$$

The non-dimensional equations of motion for the aerofoil have been derived in many textbooks of aero-elasticity, see for example Fung [26], and can be written as

$$\xi'' + x_\alpha \alpha'' + 2\zeta_\xi \frac{\bar{\omega}}{U^*} \xi' + \left(\frac{\bar{\omega}}{U^*}\right)^2 \xi = -\frac{1}{\pi\mu} C_L(S) \qquad (2)$$

$$\frac{x_\alpha}{r_\alpha^2} \xi'' + \alpha'' + 2\frac{\zeta_\alpha}{U^*} \alpha' + \left(\frac{1}{U^*}\right)^2 (\alpha - \theta_0) = \frac{(0.5 + a_h)}{\pi\mu r_\alpha^2} C_N(S) + \frac{2}{\pi\mu r_\alpha^2} (C_M(S) + C_{M0}) \qquad (3)$$

where the geometric angle of incidence, $\alpha$, is measured in radians and the plunge displacement is represented by the non-dimensional quantity, $\xi = h/b$. The symbol $(')$ denotes differentiation with respect to the non-dimensional time $S = tU/b$. The offset angle $\theta_0$ is defined as the angle wherein the spring restoring moment equals zero and $C_{M0}$ denotes the static moment coefficient at zero incidence [23-25]. We highlight the coefficient $C_{M0}$ is used to break the mathematical symmetry of a symmetric dynamical system (e.g. the case in section 4.1). The remaining parameters in equations (2-3) are: the non-dimensional flow velocity $U^*$; the aerofoil radius of gyration $r_\alpha$; the non-dimensional mass ratio $\mu$; the non-dimensional damping coefficients in plunge $\zeta_\xi$ and in pitch $\zeta_\alpha$; and the ratio of natural frequencies $\bar{\omega}$.

In the LB model the generic aerodynamic coefficient $C_i$ is assumed to be function of aerofoil kinematics of the form $C_i = C_i(\alpha, \alpha', \alpha'', \xi', \xi'')$ as explained in the next section.

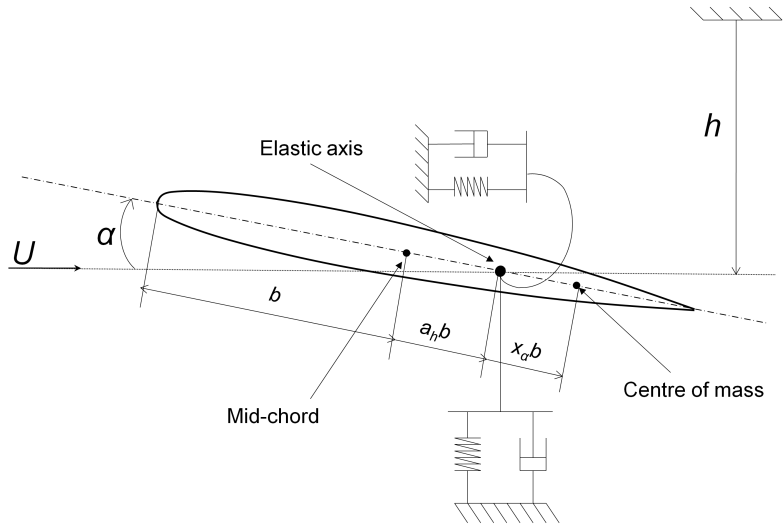

**Figure 2**: Notation used to describe the mechanical model of a typical aerofoil section.

## 3. – THE LEISHMAN-BEDDOES MODEL

This section briefly describes the state-space form of the LB model as presented in [12, 13, 27]. The main advantage of using the state-space form is that it can be appended to the structural model to construct a system of ODEs representing a dynamical system including both structural and aerodynamic states. Only the most relevant features of the model will be presented here, more details can be found in references [23,24, 25, 28].

The dynamic stall model is represented by a set of ODEs of the form

$$x'_{LB} = f_{LB}(x_{LB}, \alpha_e, q), \qquad (4)$$

where $x'_{LB} = dx_{LB}/dS$. The state vector is $x_{LB} = \{x_{LB,1}, x_{LB,2}, ..., x_{LB,n}\}^T$, and $n$ is the number of aerodynamic states, 12 in the LB model, while $f_{LB}$ is the RHS of the LB model. The inputs to the system are the effective pitch angle, $\alpha_e$, and the non-dimensional pitch rate, $q$. Appropriate definitions of $\alpha_e$ and $q$, which take into account the effects of plunging, are given and further discussed below.

The aerodynamic lift and pitching moment coefficients are given by

$$C_L = C_L(x_{LB}, \alpha_e, q), \qquad C_M = C_M(x_{LB}, \alpha_e, q) \qquad (5)$$

The dynamical system is composed of two terms: the equations of motion of the airfoil (2) and (3), and the equations of the dynamic stall model (4). The motion of the airfoil will determine the quantities $\alpha_e$ and $q$, which are given as inputs to the dynamic stall model. At the same time the dynamic stall model will provide the aerodynamic forces required by the mechanics of the airfoil. Using the quasi-steady approach [28], the effective angle of incidence is given by [24, 28]

$$\alpha_e = \alpha_e(\alpha, \alpha', \xi'). \qquad (6)$$

The LB model also requires the pitch rate, $q$, which should be calculated as twice the derivative of the above expression, and therefore assumes the form

$$q = q(\alpha, \alpha', \alpha'', \xi', \xi''). \qquad (7)$$

The non-dimensional equations of motion (2) and (3) can be transformed into four first order ODEs by writing

$$\hat{x}' = \hat{f}(x_{LB}, \hat{x}) \qquad (8)$$

where $\hat{x} = \{\alpha, \alpha', \xi, \xi'\}^T = \{\hat{x}_{13}, \hat{x}_{14}, \hat{x}_{15}, \hat{x}_{16}\}^T$ are the states representing the motion of the aerofoil and $\hat{f}$ is consequently the RHS of the structural model. Finally the system of four ODEs (8), for states 13 to 16, must be appended to the LB model state-space formulation as defined in equation (4) to complete the aero-elastic system of 16 first order ODEs. The state variable vector and the RHS of the aero-elastic system are denoted by $x$ and $f$ respectively. Hence the aero-elastic system is of the form

$$x' = f(x, \alpha_e, q) \qquad (9)$$

The presence of a term such as $C_L(\alpha_e, q)$ in eq. 2 makes the system of differential equation (9) implicit and therefore more difficult and expensive to solve numerically. However, we will neglect the acceleration terms, imposing $\xi''=\alpha''=0$, in the expression for the pitch rate, i.e. we will assume $q=q(\alpha,\alpha',\xi')$. The introduction of this simplification appears to be common practice, see for instance [24, 29], and we have checked that this has a negligible impact on the dynamical behaviour of the system when the parameters assume the values given later.

The differential system (9) is integrated with the adaptive Runge-Kutta-Fehlberg routine ode45 of the Matlab suite [30] equipped with an event detection algorithm. As it is apparent from the equations of the LB model given in the appendix B, some coefficients of the equations of motion, or even the form of one or more equations, vary when certain algebraic equalities are satisfied. Non-smooth differential systems have often to be integrated with routines equipped with event detection algorithms [23]. The role of an event detection algorithm is to locate the values of all state variables where a discontinuity along a trajectory occurs. At that point, the system of equations has to be modified and the numerical integration restarted with the initial conditions provided by the event detector.

### 3.1. – ON THE NON-SMOOTHNESS OF THE LB MODEL

The state-space LB model (4) produces as outputs the coefficients $C_N$, $C_C$ and $C_M$. The model consists of three distinct parts: attached flow, trailing edge separated flow and dynamic stall (vortex induced airloads). Typically the aerodynamic coefficients $C_i$; $i=N, C, M,$ are calculated as sum of three components: $C_i = C_i^P + C_i^f + C_i^v$, where the superscripts $P$, $f$ and $v$ refer to the attached flow terms, the trailing edge separation and the vortex induced loads, respectively.

#### 3.1.1 Attached flow

The unsteady attached flow is modelled using unsteady thin aerofoil theory (based on the potential function) and its behaviour is represented by a linear system of equations of the form

$$\{x_1' \quad \mathrm{K} \quad x_8'\}^T = \mathbf{A}\{x_1 \quad \mathrm{K} \quad x_8\}^T + \mathbf{B}\{\alpha_e \quad q\}^T \tag{10}$$

where $\mathbf{A}$ is an 8×8 constant matrix, $\mathbf{B}$ is an 8×2 constant matrix. The attached flow parts of $C_N$ and $C_M$ are therefore defined as linear functions of $x_1$-$x_8$

$$\{C_N^P \quad C_M^P\}^T = \mathbf{C}\{x_1 \quad \mathrm{K} \quad x_8\}^T + \mathbf{D}\{\alpha_e \quad q\}^T \tag{11}$$

where $\mathbf{C}$ is an 8x2 constant matrix, $\mathbf{D}$ is a 4x4 constant matrix. The chord force coefficient in unsteady attached flow equals zero, $C_C^P = 0$. Matrices $\mathbf{A}$, $\mathbf{B}$, $\mathbf{C}$ and $\mathbf{D}$ are given in reference [23] for the case of a NACA 0012 profile.

#### 3.1.2 Trailing edge separation

The trailing edge separation parts of $C_N$, $C_C$, $C_M$ are defined as non-linear functions of $x_9$-$x_{11}$ which are obtained as the solution of the system of delay ODEs

$$\begin{Bmatrix} x_9' \\ x_{10}' \\ x_{11}' \end{Bmatrix} = \begin{Bmatrix} (C_N^P - x_9)/T_P \\ [f(x_9/C_{N\alpha}, \alpha_1) - x_{10}]/T_f \\ [f(\alpha_e, \alpha_1) - x_{11}]/(0.63 T_{f0}) \end{Bmatrix} \tag{12}$$

The function $f$ is non-smooth and is given by the modified-Kirchhoff theory in appendix A. The dynamic stall onset takes place as the state variable $x_9$ reaches the value of the critical normal force coefficient, $C_{N1}$. Thereby, when $|x_9|=C_{N1}$ and $|x_9|$ is increasing, a time-like variable $\tau_v$ starts from 0 and the values of $T_f$ vary discontinuously as indicated in table 1.

Table 1: Variation of parameter $T_f$ according to the LB model [23].

|  | $0 \leq \tau_v \leq T_{vl}$ | $T_{vl} < \tau_v \leq 2T_{vl}$ | $2T_{vl} < \tau_v$ |
|---|---|---|---|
| $\alpha_e \cdot \alpha_e' \geq 0$ | $T_f = 3T_{f0}$ | $T_f = T_{f0}/3$ | $T_f = 4T_{f0}$ |
| $\alpha_e \cdot \alpha_e' < 0$ | $T_f = 0.5T_{f0}$ | $T_f = 0.5T_{f0}$ | $T_f = 4T_{f0}$ |

The parameter $\alpha_1$ does not depend on $\tau_v$ and is given by:

$$\alpha_1 = \begin{cases} \alpha_{10} & \text{if} \quad \alpha_e \alpha_e' \geq 0 \\ \alpha_{10} - (1-x_{10})^{0.25} \delta_{\alpha 1} & \text{if} \quad \alpha_e \alpha_e' < 0 \end{cases} \tag{13}$$

On the other hand, when $|x_9|=C_{N1}$ and $|x_9|$ is decreasing, then reattachment begins and $T_f$ varies as follows:

$$T_f = \begin{cases} T_{f0} & \text{if } x_{10} \geq 0.7 \\ 2T_{f0} & \text{if } x_{10} < 0.7 \end{cases} \quad \text{and} \quad \alpha_1 = \alpha_{10} \tag{14}$$

where $\alpha_{10}$ is the experimental static stall angle, while $T_P$, $T_{f0}$ and $\delta_{\alpha 1}$ are other experimental parameters [23]. The corresponding aerodynamic coefficients are predicted by the modified-Kirchhoff theory (see appendix A).

### 3.1.3 Vortex induced airloads
The vortex induced contribution is defined by means of a delay equation of the form

$$x'_{12} = \begin{cases} c'_v - x_{12}/T_v & \text{if} \quad \alpha_e c'_v \geq 0 \text{ and } 0 < \tau_v < 2T_{vl} \\ -x_{12}/T_v & \text{otherwise} \end{cases} \tag{15}$$

where $T_v$ is affected by jumps similar to those of $T_f$. The parameter $c_v$ represents the strength of vortex induced normal force. The corresponding aerodynamic coefficients are

$$C_N^v = x_{12}, \quad C_C^v = 0, \quad C_M^v = C_M^v(x_{12}, \tau_v). \tag{16}$$

### 3.2 - Location and stability of fixed points
Since the size of the state-space aero-elastic model is 16, the co-ordinates of a fixed point of the system are 16 as well. From a physical standpoint, we focus our attention on the 13$^{th}$ co-ordinate of the fixed point, namely the geometric angle of incidence $x_{13}$. In order to calculate it we impose both the derivatives $x'$ and the velocity terms $q$, $x_{14}$, $x_{16}$ equal to zero, because at the fixed point the system is static by definition. We emphasize at a fixed point the effective angle of incidence $\alpha_e$ equals the geometric angle of incidence $x_{13}$ and, given that $\alpha' = \alpha_e' = 0$, the parameter $\alpha_1$ equals the static stall angle $\alpha_{10}$ (eq. 13). Furthermore, in static flows the physical condition $x_{13} = \alpha_{10}$ involves the equalities $x_{10} = 0.7$ and $x_{13} = x_9/C_{N\alpha}^S$. We calculate the position of a fixed point just solving the algebraic system $f(x)=0$, where $f$ is the RHS of the aero-elastic system (see appendix B). The 13$^{th}$ equation of this algebraic system, which provides the geometric angle of incidence $x_{13}$, is uncoupled from the rest. Hence, we have to solve just one non-linear algebraic equation:

$$\left(0.5 U^{*-2}\pi\mu r_\alpha^2\right)(x_{13} - \theta_0) - 0.5(0.5 + a_h)C_N^S - C_M^S = 0 \tag{17}$$

where the expressions of the static coefficients $C_M^S$ and $C_N^S$ are predicted by using the static-Kirchhoff theory [31] (see appendix A). The other co-ordinates of the fixed point depend only on the geometric angle of incidence $x_{13}$ [25, 28]. Stability of fixed points can be discussed as usual by computing the eigenvalues of the Jacobian matrix:

$$J = \frac{\partial f}{\partial x}\bigg|_x \tag{18}$$

The Jacobian matrix is a square matrix 16x16 and contains the first order partial derivatives of $f$ evaluated at the fixed point $x$. In index notation, the components of the Jacobian matrix are represented by $J_{ij} = \partial f_i/\partial x_j$ where $i,j=1,2,...,16$. The stability of the aero-elastic system is determined by analysing the eigenvalues $\lambda_i$ of the Jacobian matrix. If all real parts are negative the system is stable. However, if the real part of one eigenvalue is positive the system is unstable. By definition, at the Hopf bifurcation a pair of conjugate eigenvalues will cross the imaginary axis.

### 3.3 - Equations identifying the discontinuity of interest

As we'll see in section 4, the largest real part of the eigenvalues may cross the imaginary axis discontinuously. In this work, this jump is due to the following components of the Jacobian matrix:

$$\frac{\partial f_{10}}{\partial x_9} = \begin{cases} -\text{sign}(x_9) \cdot 0.3 / (T_f C_{N\alpha}^S S_1) \cdot \exp\left[(|x_9 / C_{N\alpha}^S| - \alpha_{10}) / S_1\right] & |x_9 / C_{N\alpha}^S| \leq \alpha_{10} \\ -\text{sign}(x_9) \cdot 0.66 / (T_f C_{N\alpha}^S S_2) \cdot \exp\left[(-|x_9 / C_{N\alpha}^S| + \alpha_{10}) / S_2\right] & |x_9 / C_{N\alpha}^S| > \alpha_{10} \end{cases} \quad (19)$$

$$\frac{\partial f_{12}}{\partial x_{13}} = \begin{cases} -\text{sign}(x_{13}) \cdot 0.3 / (0.63 T_{f0} C_{N\alpha}^S S_1) \cdot \exp\left[(|x_{13}| - \alpha_{10}) / S_1\right] & |x_{13}| \leq \alpha_{10} \\ -\text{sign}(x_{13}) \cdot 0.66 / (0.63 T_{f0} C_{N\alpha}^S S_2) \cdot \exp\left[(-|x_{13}| + \alpha_{10}) / S_2\right] & |x_{13}| > \alpha_{10} \end{cases} \quad (20)$$

where $f_n$ is the n-th equation of the RHS of the aero-elastic system at the fixed point. The discontinuity locus is given by the condition $x_{13}=\alpha_{10}$, that defines two different zones of the phase space in which the model is defined by two different sets of equations depending on the value of the angle of incidence $x_{13}$, evaluated at the fixed point. Although there are other discontinuities conditions within the LB model (e.g. $x_9=C_{N1}$), we are not interested in them because the discontinuous behaviour analysed in section 4 only refers to the above equations (19-20). From now on, we'll refer the set of equations valid for $|x_{13}| = |x_9 / C_{N\alpha}^S| \leq \alpha_{10}$ as the *first set of equations*, and that valid for $|x_{13}| = |x_9 / C_{N\alpha}^S| > \alpha_{10}$ as the *second set of equations*.

The static stall angle $\alpha_{10}$ is a function of the Mach number as given in table 2 taken from [23] for a profile NACA 0012 and standard air flows with Reynolds number Re=$3.6 \times 10^6$.

**Table 2**: Experimental static stall angle $\alpha_{10}$ as a function of Mach [23].

| Mach | 0.3 | 0.4 | 0.5 | 0.6 | 0.7 | 0.75 | 0.8 |
|---|---|---|---|---|---|---|---|
| $\alpha_{10}$[deg] | 14.49 | 11.87 | 9.98 | 8.07 | 5.32 | 3.23 | 0.66 |

**4. – EXAMPLES OF HOPF BIFURCATIONS**

Both aero-elastic systems presented in this section have in common the following parameters: $C_{M0}$=-0.0037; $\mu$=36.504; $\bar{\omega} = 0.5$; $r_\alpha$=0.5; $a_h$=-0.5 and $\zeta_\alpha = \zeta_\xi = 0$. The natural circular frequency in plunge is $\omega_\xi$=100 rad/s. Therefore the free parameters of the system are the Mach number $M$ and the offset angle $\theta_0$. In our calculations we used an interpolation of the experimental data (e.g. $\alpha_{10}$, $S_1$, $S_2$, $T_{f0}$, etc ...) made available by Galvanetto et al. [23] as functions of Mach.

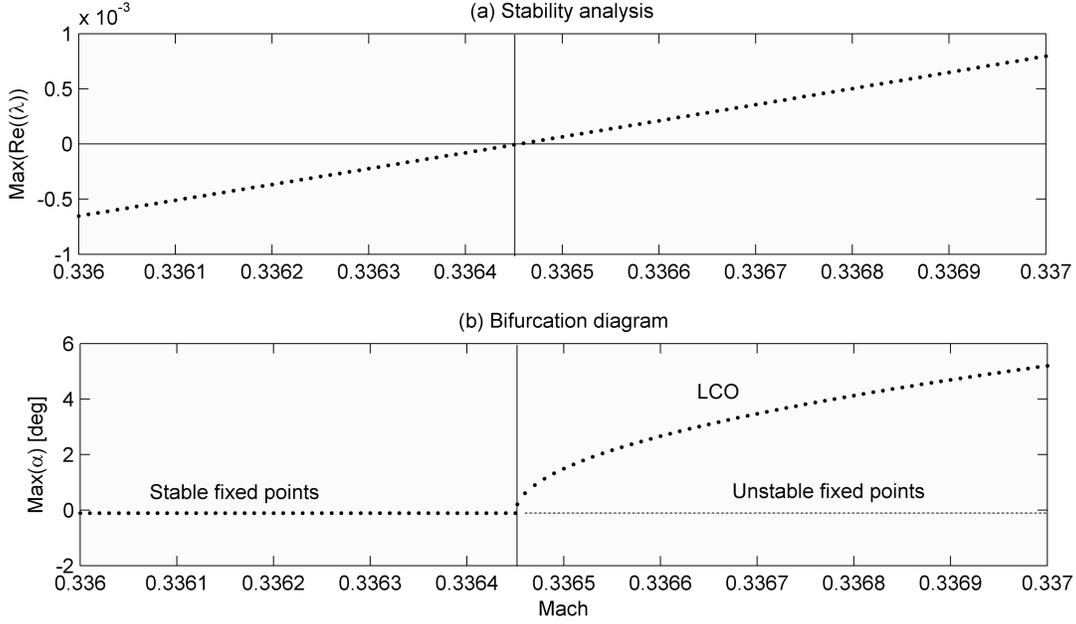

**Figure 3:** (a) Variation of the largest real part of the system eigenvalues and (b) bifurcation diagram of the relevant standard Hopf bifurcation.

### 4.1 Standard Hopf bifurcation:
We have investigated an aero-elastic system with an offset angle $\theta_0=0°$. Such a system has a fixed point which does not intersect the discontinuities of the LB model, therefore the aero-elastic system behaves like a smooth system. From linearised analysis we know that the supercritical Hopf bifurcation takes place at Mach ≈ 0.33645 (fig. 3.a). In the bifurcation diagram (fig. 3.b) we have plotted a mono-dimensional projection of the Poincaré map of the aero-elastic system. The steady solution changes topology at the critical point, where the fixed point becomes unstable and an LCO grows continuously. This kind of behaviour is typical of smooth dynamical systems.

### 4.2 Non-smooth Hopf bifurcation:
The non-smoothness of the LB model can affect the bifurcation diagram remarkably. In this section, we have investigated a case in which the value of the offset angle $\theta_0=14°$ was chosen a little lower than the one of the static stall angle $\alpha_{10}$. Thus as the Mach number increases the position of the fixed point crosses over the discontinuity line defined by the static stall angle. In fig. 4, we notice that the path of the fixed point (thick line) has a point of non-differentiability at the intersection with the discontinuity line (thin line). As the fixed point crosses the discontinuity line the Jacobian matrix undergoes a jump (eq. 19-20) and the same happens to its eigenvalues in a way that the largest real part of the eigenvalues, which is negative on one side of the discontinuity line becomes positive on the other (fig. 5.a). This discontinuous variation of the eigenvalues triggers a sudden loss of stability of the fixed point at Mach ≈ 0.30155 where a discontinuous supercritical Hopf bifurcation takes place. In the bifurcation diagram (fig. 5.b) we have plotted the maximum angle of incidence of the steady solution, obtained by integration in the non-dimensional time domain. We observe that the solution changes topology at the critical point abruptly, passing from a fixed point to a finite amplitude periodic attractor (LCO) as Mach increases. This type of bifurcation is a peculiar characteristic of non-smooth dynamical systems: no smooth system can present such a discontinuous behaviour. To fully grasp this distinctive feature of non-smooth systems we refer to figure 6. For values of Mach just below the bifurcation value the first set of equation holds and the fixed point is stable.

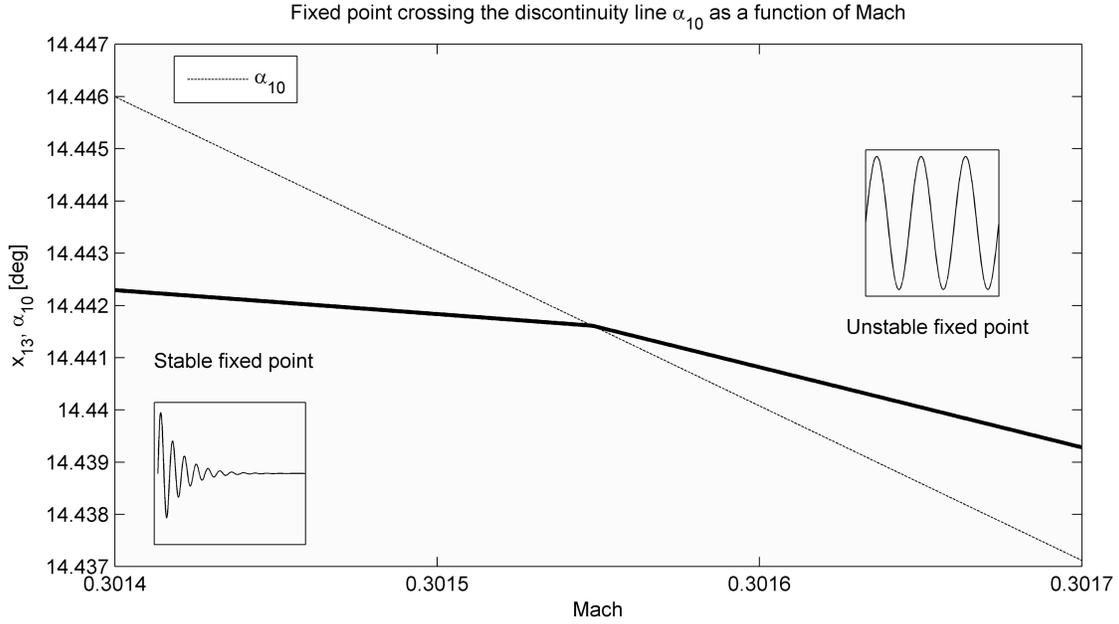

**Figure 4:** Co-ordinate $x_{13}$ of the fixed point (thick line) and value of the static stall angle $\alpha_{10}$ (thin line) as functions of Mach. If $x_{13}$ at the fixed point is less than $\alpha_{10}$ the fixed point is stable, otherwise unstable. The first set of equations is used for $x_{13} \leq \alpha_{10}$ and the second for $x_{13} > \alpha_{10}$ for that reason the thin line is called discontinuity line.

The solution of the time integration with initial conditions close to the fixed point is a spiral trajectory converging to the stable fixed point (fig. 6.a). Conversely, for values of Mach just above the bifurcation value the second set of equations holds and the fixed point is unstable. The trajectory originating from a point close to the unstable fixed point converges to an LCO (fig. 6.b). The transition from fixed point to LCO and vice versa will not occur gradually but in a discontinuous way.

The co-ordinate $x_6$ of the fixed point in fig. 6(a) was calculated through the relation $x_6 = -x_{13}/a_{66}$ taking as an input the variable $x_{13}$ shown in fig. 4. This relation is produced applying the definition of fixed point to the eq. B.1 (appendix B, for i =6), substituting the 13$^{th}$ component of the fixed point $x_{13}$ to $\alpha$ and imposing the static condition $x_6'=0$. In a similar way the thin line in figure 6(a) is given by the equation $x_6 = -\alpha_{10}/a_{66}$ and therefore represents the discontinuity locus in the two-dimensional projection of the phase space.

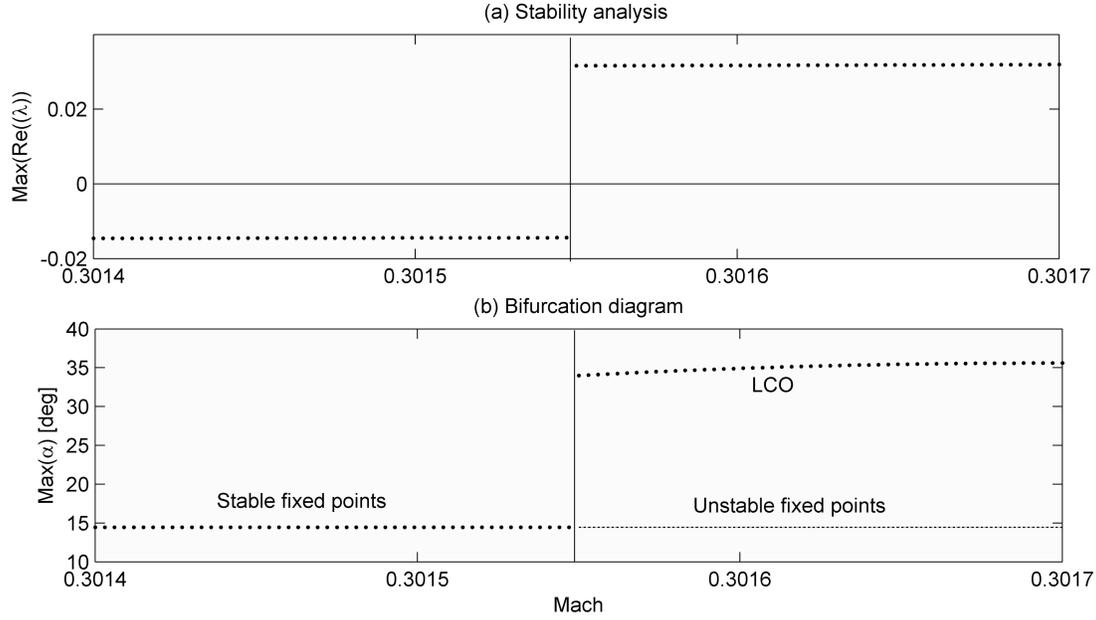

**Figure 5:** (a) Variation of the largest real part of the system eigenvalues. (b) Bifurcation diagram of the relevant non-smooth Hopf bifurcation.

## 5 CONCLUSIONS

The present paper investigates possible interactions between the discontinuous definition of the equations of the Leishman and Beddoes (LB) dynamic stall model and the main Hopf bifurcation of the aero-elastic system to which they are appended. The free-stream Mach number, $M$, has been used as the bifurcation parameter for all numerical computations. Provided that the free-stream speed of sound remains constant, the free-stream Mach number at the Hopf bifurcation can also be recognised as the flutter speed, $U_c$. Often aero-elastic systems, which are stable for low values of the undisturbed airflow speed $U$, lose their stability at a critical value of $U_c$, at which they start oscillating. If a non-linear model is adopted at this main Hopf bifurcation a limit cycle is generated. In smooth systems the amplitude of the limit cycle, infinitesimal at the bifurcation point, grows continuously for increasing values of $U$. In a two-degree-of-freedom aerofoil equipped with the LB model, such a standard Hopf bifurcation is clearly possible, however even another type of non-smooth Hopf bifurcation is possible, in particular if the undisturbed equilibrium position of the aerofoil is close to the static stall angle. In the non-smooth Hopf bifurcation the amplitude of the limit cycle generated at the bifurcation is immediately finite. It is apparent that the non-smooth Hopf bifurcation is an artefact of the model which does not describe any type of natural phenomenon. Bifurcation diagrams showed that changes in the steady states may be caused by the interaction between the discontinuity boundaries and the trajectories. Hence, smoothing the discontinuities could modify significantly the bifurcation diagrams. This could be performed by replacing the step changes in parameter values and the sudden switching of ODEs, occurring at the discontinuity boundaries, by smooth transitions.

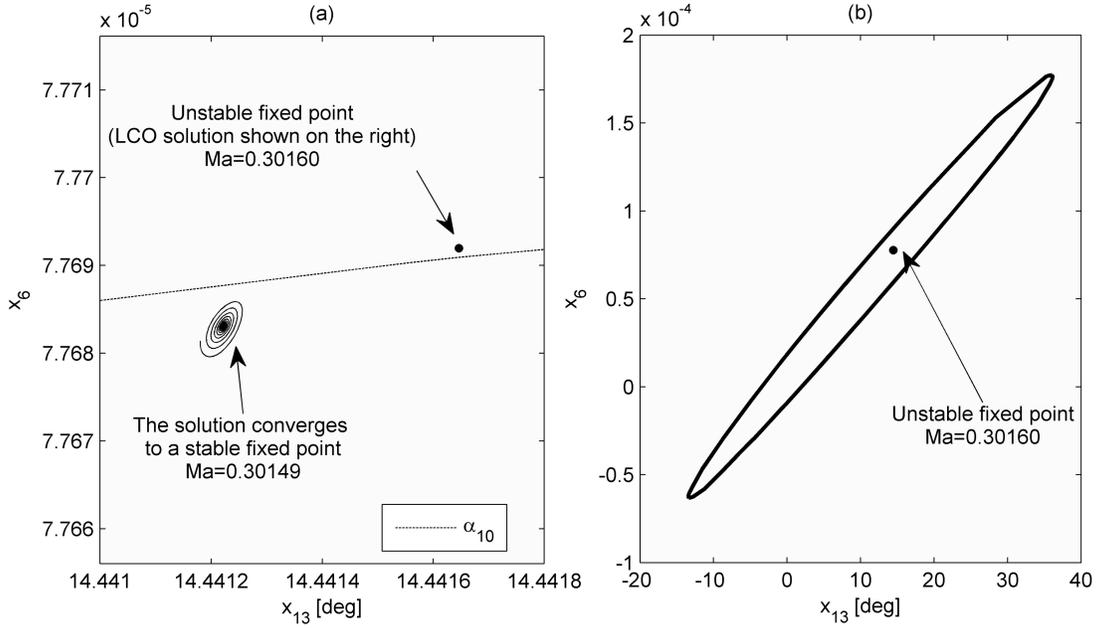

**Figure 6:** Projection of trajectories in the plane $x_6$-$x_{13}$: (a) the spiral trajectory converges to the stable fixed point, (b) the stable limit cycle born around the unstable fixed point.

**ACKNOWLEDGMENTS**

The authors would like to thank Dr. Chawin Chantarasenawong (Department of Mechanical Engineering, King Mongkut's University of Technology, Thonburi, Bangkok, Thailand) for his help.

**APPENDIX**
**A- MODIFIED-KIRCHHOFF THEORY ADAPTED TO DYNAMIC CONDITIONS**

$$f(\alpha, \alpha_1) = \begin{cases} 1 - 0.3 \exp\left[(|\alpha| - \alpha_1)/S_1\right] & |\alpha| \leq \alpha_1 \\ 0.04 + 0.66 \exp\left[(-|\alpha| + \alpha_1)/S_2\right] & |\alpha| > \alpha_1 \end{cases} \quad (A.1)$$

$$C_N = 0.25 C_N^C \left(1 + \sqrt{x_{10}}\right)^2; \qquad C_C = 0.97 C_{N\alpha}^S \left(C_N^C / C_{N\alpha}^S\right)^2 \sqrt{x_{10}} \quad (A.2)$$

$$C_M = C_N \left[K_0 + K_1(1 - \hat{x}) + K_2 \sin(\pi \hat{x}^2)\right]; \quad \hat{x} = \begin{cases} x_{10} & x_{10} > x_{12} \\ x_{12} & x_{10} \leq x_{12} \end{cases} \quad (A.3)$$

where $C_N^C$ is the circulatory part of the normal force coefficient and $C_{N\alpha}^S$ is the static normal force coefficient slope. Substituting $f$ to $\hat{x}$ and to $x_{10}$ we return to the classical static-Kirchhoff theory [31].

**B- RIGHT-HAND SIDE OF THE AERO-ELASTIC SYSTEM**
In this appendix we render explicit the 16 components of the RHS of the aero-elastic system $f$, previously represented in the general mathematical form as the RHS of an ODEs system $x'=f(x)$.

$$f_i = \begin{cases} (b/U)(a_{ii} x_i + \alpha + 0.5 q) & i = 1, 2 \\ (b/U)(a_{ii} x_i + \alpha) & i = 3, 5, 6 \\ (b/U)(a_{ii} x_i + q) & i = 4, 7, 8 \end{cases} \quad (B.1)$$

$$f_9 = (1/T_P)\left[c_{11}x_1 + c_{12}x_2 + c_{13}x_3 + c_{14}x_4 + (4/M)\alpha + (1/M)q - x_9\right] \tag{B.2}$$

$$f_{10} = \begin{cases} 1/T_f \cdot \left\{1 - 0.3\exp\left[\left(\left|x_9/C_{N\alpha}^S\right| - \alpha_1\right)/S_1\right] - x_{10}\right\} & \left|x_9/C_{N\alpha}^S\right| \le \alpha_1 \\ 1/T_f \cdot \left\{0.04 + 0.66\exp\left[\left(-\left|x_9/C_{N\alpha}^S\right| + \alpha_1\right)/S_2\right] - x_{10}\right\} & \left|x_9/C_{N\alpha}^S\right| > \alpha_1 \end{cases} \tag{B.3}$$

$$f_{11} = -x_{11}/T_v \tag{B.4}$$

$$f_{12} = \begin{cases} 1/(0.63T_{f0}) \cdot \left\{1 - 0.3\exp\left[\left(\left|x_{13}\right| - \alpha_1\right)/S_1\right] - x_{12}\right\} & \left|x_{13}\right| \le \alpha_1 \\ 1/(0.63T_{f0}) \cdot \left\{0.04 + 0.66\exp\left[\left(-\left|x_{13}\right| + \alpha_1\right)/S_2\right] - x_{12}\right\} & \left|x_{13}\right| > \alpha_1 \end{cases} \tag{B.5}$$

$$f_{13} = x_{14} \tag{B.6}$$

$$f_{14} = \left(x_\alpha^2/r_\alpha^2 - 1\right)^{-1}\left[U^{*-2}(x_{13} - \theta_0) + \left(2\zeta_\alpha U^{*-1}\right)x_{14} - \left(U^{*-2}r_\alpha^{-2}x_\alpha\bar{\omega}^2\right)x_{15} + \right.$$
$$\left. -\left(2\zeta_\xi U^{*-1}r_\alpha^{-2}x_\alpha\bar{\omega}\right)x_{16} - \left(\pi\mu r_\alpha^2\right)^{-1}\left(x_\alpha C_L + (0.5 + a_h)C_N + 2(C_M + C_{M0})\right)\right] \tag{B.7}$$

$$f_{15} = x_{16} \tag{B.8}$$

$$f_{16} = x_\alpha\left(x_\alpha^2/r_\alpha^2 - 1\right)^{-1}\left[-U^{*-2}(x_{13} - \theta_0) - \left(2\zeta_\alpha U^{*-1}\right)x_{14} + \left(x_\alpha^{-1}U^{*-2}\bar{\omega}^2\right)x_{15} + \right.$$
$$\left. +\left(2\zeta_\xi\bar{\omega}x_\alpha^{-1}U^{*-1}\right)x_{16} + \left(\pi\mu r_\alpha^2\right)^{-1}\left(x_\alpha^{-1}r_\alpha^2 C_L + (0.5 + a_h)C_N + 2(C_M + C_{M0})\right)\right] \tag{B.9}$$

where $M$ is the Mach number. The coefficients $a_{ii}$ are diagonal components of the matrix $\mathbf{A}$ (eq. 10), while $c_{11}, c_{12}, c_{13}$ and $c_{14}$ are four components of the matrix $\mathbf{C}$ (eq. 11).

## REFERENCES

[1] McCroskey W. J. Unsteady Airfoils. Annual Review of Fluid Mechanics, 14:285–311, 1982.

[2] Ericsson L. E. and Reding J. P. Fluid Mechanics of Dynamic Stall Part 1. Unsteady Flow Concepts. Journal of Fluids and Structures, 2:1–33, 1988.

[3] Ericsson L. E. and Reding J. P. Fluid Mechanics of Dynamic Stall Part 2. Prediction of Full Scale Characteristics. Journal of Fluids and Structures, 2:113–143, 1988.

[4] Leishman J. G. Principles of Helicopter Aerodynamics. Cambridge University Press, Second edition, 2006.

[5] Petot D. Modélisation du Décrochage Dynamique. La Recherche Aérospatiale, 5:60–72, 1989.

[6] Truong K. V. and Costes J. J. Oscillatory Behaviour of Helicopter Rotor Airloads in the Blade Stall Regime. Journal of Aircraft, 32(5):1148–1149, 1995.

[7] Gormont R. E. A Mathematical Model of Unsteady Aerodynamics and Radial Flow for Application to Helicopter Rotors. USAAVLABS, TR 72-67, May 1973.

[8] Johnson W. The Response and Airloading of Helicopter Rotor Blades Due to Dynamic Stall. Massachusetts Institute of Technology, ASRL TR 130-1, May 1970.

[9] Gangwani S. T. Prediction of Dynamic Stall and Unsteady Airloads For Rotor Blades. Journal of the American Helicopter Society, 27:57–64, 1982.

[10] Leishman J. G. and Beddoes T. S. A Generalised Model for Airfoil Unsteady Aerodynamic Behaviour and Dynamic Stall Using the Indicial Method. In Proceedings of the 42nd Annual Forum, Washington DC, 1986. American Helicopter Society.

[11] Leishman J. G. and Beddoes T. S. A Semi-Empirical Model for Dynamic Stall. Journal of the American Helicopter Society, 34:3–17, 1989.

[12] Leishman J. G. and Nguyen K. Q. State-Space Representation of Unsteady Airfoil Behaviour. AIAA Journal, 28(5):836–844, 1988.

[13] Leishman J. G. and Crouse G. L. Jr. State-Space Model for Unsteady Airfoil Behaviour and Dynamic Stall. Proceedings of the AIAA/AHS/ASME Structural Dynamics and Materials Conference, Mobile, Alabama, April 1989. AIAA paper 89-1319.

[14] Hansen M. H., Gaunaa M., and Madsen H. A. A Beddoes-Leishman Type Dynamic Stall Model in State-Space and Indicial Formulation. Risø National Laboratory, Roskilde, Denmark, June 2004.

[15] Larsen J. W., Nielsen S. R. K., and Krenk S. Dynamic Stall Model for Wind Turbine Airfoils. Journal of Fluids and Structures 23 (2007), 959-982.

[16] Leine R. I. and Nijmeijer H. Dynamics and Bifurcations in Non-Smooth Mechanical Systems, volume 18 of Lecture notes in Applied and Computational Mechanics. Springer-Verlag, 2004.

[17] Di Bernardo M., Budd C.J., Champneys A.R., Kowalczyk P. Piecewise-smooth dynamical systems: theory and applications, Springer-Verlag (Applied Mathematics series no. 163), November 2007.

[18] Galvanetto U., Bishop, SR, Stick-slip vibrations of a 2-degree-of-freedom geophysical model, Int.. J. Mec. Sciences, 36: 8, 683-698, 1994.

[19] Galvanetto U. Numerical computation of Lyapunov exponents in discontinuous maps implicitly defined, Comp. Phys. Comm., 131: 1-2, 1-9, 2000.

[20] Galvanetto U, Non-linear dynamics of multiple friction oscillators, Comp. Meth. Appl. Mech. Eng., 178: 3-4, 291-306, 1999.

[21] Price S. J. and Keleris J. P. Non-linear Dynamics of an Airfoil Forced to Oscillate in Dynamic Stall. Journal of Sound and Vibration, 194(2):265–283, 1996.

[22] Li X. G. and Fleeter S. Dynamic Stall Generated Airfoil Oscillations. International Journal of Turbomachinery and Jet Engines, 20:217–233, 2003.

[23] Galvanetto U., Peirò J., Chantarasenawong C. An Assessment of some Effects of the Non-smoothness of the Leishman-Beddoes Dynamic Stall Model on the Nonlinear dynamics of a Typical Aerofoil Section, Journal of Fluids and Structures 24 (2008), 151-163.

[24] Peirò J., Galvanetto U., Chantarasenawong C. Assessment of Added Mass Effects on Flutter Boundaries using the Leishman-Beddoes Dynamic Stall Model, Journal of Fluids and Structures 26 (2010), 814-840.

[25] Chantarasenawong C. Nonlinear Aero-elastic Behaviour of airfoils Under Dynamic Stall, PhD thesis, Imperial College London, 2007.

[26] Fung Y. C. An Introduction to the Theory of Aero-elasticity. Dover, New York, 1993.

[27] Crouse G. L. Jr. and Leishman J. G. Transonic Aero-elasticity Analysis Using State-Space Unsteady Aerodynamic Modeling. Journal of Aircraft, 29(1):153–160, 1992.

[28] Magri L., Non-linear Aero-elastic Analysis of Stall Flutter modelled as a Piece-wise Smooth Dynamical System, Master thesis (tesi di Laurea Magistrale) in Italian, Dip. Costruz. Trasporti, Università degli studi di Padova, 2011.

[29] Mahajan A. J., Kaza K. R. V., and Dowell E. H. Semi-Empirical Model for Prediction of Unsteady Forces on an Airfoil With Application to Flutter. Journal of Fluids and Structures, 7:87–103, 1993.

[30] Shampine L. F. and Reichelt M. W. The MATLAB ODE suite. SIAM Journal on Scientific Computing, 18(1):1–22, 1997.

[31] Twhaites B. Incompressible aerodynamics, Oxford University Press, 1960.